\documentclass[acmtog,anonymous]{acmart}
\preto{\absreactkeywords}{\nolinenumbers}
\acmSubmissionID{1234}

\usepackage{booktabs} 
\let\nofiles\relax
\usepackage[ruled]{algorithm2e} 

\SetAlFnt{\small}
\SetAlCapFnt{\small}
\SetAlCapNameFnt{\small}
\SetAlCapHSkip{0pt}

\acmJournal{TOG}

\begin{document}
\title{A Gpu-based solution for large-scale skeletal animation simulation}

\begin{abstract}
Skeletal animations of large-scale characters are widely used in video games. However, with a large number of characters are involved, relying on the CPU to calculate skeletal animations leads to significant performance problems. There are two main types of traditional GPU- based solutions. One is referred to as pre-baked animation texture technology. The problem with this solution is that it can only play animations from the pre-baked animation. It is impossible to perform interpolation, blending and other calculations on the animation, which affects the quality of the animations. The other solution is referred to as dedicated processing with a simple skeleton hierarchy (the number of skeleton levels < 64). This option does not need to simulate and bake animation data in advance. However, performance is dramatically impaired when processing complex skeletons with too many skeleton levels (such as fluttering clothing, soft plants, dragon-like creatures, etc.). In order to solve these issues, we developed a parallel prefix tree update solution to optimize the animation update process of complex skeletons with too many levels, and combined traditional solutions to implement a GPU-based skeletal animation solution. This solution does not need to simulate and bake animation results. In addition, the performance is superior to traditional solutions for complex skeletons with too many levels. Our work can provide a new option for optimizing the performance of large-scale skeletal animation simulations, providing GPU-based skeletal animations a wider range of application scenarios.

\end{abstract}

%
%

%
%


\maketitle

\let\nofiles\relax
\section{Introduction}
Skeletal animation simulation of group characters is widely used in modern large-scale $3D$ game development. Games such as "Mount And Blade", "Final Fantasy $14$" and "Dyson Sphere Program" often need to render thousands of characters with skeletal animation in the same scene, and each character requires smooth and realistic animation performance. When facing this level of load, traditional CPU-based skeletal animation simulation solutions will reduce the game's frame rate, affecting the player's gaming experience. Therefore, it is crucial for modern games to use GPU improving the simulation performance of game animation.

Common skeletal animation simulation usually requires the following steps(Figure~\ref{fig:one} ):
\begin{figure}[htbp]
  \includegraphics[width=0.5\textwidth]{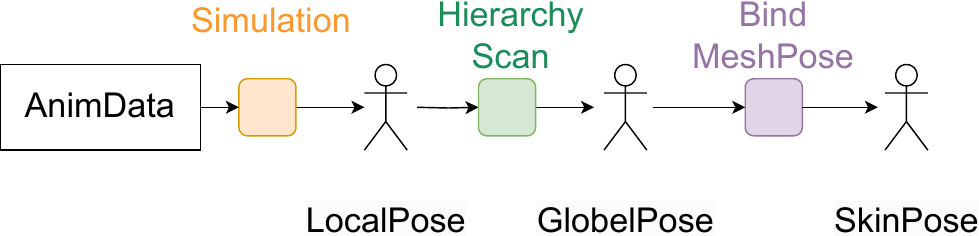}
  \caption{The processing of animation simulation}
  \label{fig:one}
\end{figure}
\begin{enumerate}
\item Simulation.
  Sample animation data and generate local pose in local space based on game and animation simulation logic.
\item Hierarchy-Scan.
  Transform local space pose to model space pose(globel pose).
\item Bind MeshPose. 
  Combine model space pose and bindpose to skinpose.
\end{enumerate}

Among them, the first step (Simulation) and the third step (Bind MeshPose) are easier to implement on the GPU because the calculation of each skeleton joint does not rely on other bones. However, the second step (Hierarchy Scan) is more difficult to implement efficiently on the GPU because the calculation of each skeleton joint depends on its parent node. Therefore, the focus of this article is mainly on how to efficiently implement Hierarch-Scan on the GPU.

\section{PREVIOUS WORK}
Early research on accelerated skeletal animation simulation based on GPUs focused on using pre-baking solutions. \cite{dudash2007skinned}The skeletal animation data is simulated in the CPU in advance, the result is baked into the animation texture, and the GPU replaces the animation simulation by sampling the texture. This work avoids the implementation of the HierarchyScan process in the GPU. However, because the animation data is baked in advance, the character's animation cannot be blended according to the game logic and animation state machine. The effect of animation interpolation is also poor.

 As an optimization\cite{shopf2008march},the bone hierarchy is flattened in the preprocessing stage. This solution can perform simple animation blending and interpolation when the bone hierarchy is simple. However, since the original structure of the skeleton is destroyed, it is less effective for skeletons with complex hierarchies.

Another solution proposes to allocate a thread to each joint of the skeleton in the GPU\cite{Gateau2012monte}. First, a compute shader is used to complete the animation interpolation sampling and blending simulation for each joint in local space. Then each joint will recursively multiply all its parent nodes until the root node completes the Hierarchy-Scan stage(Algorithm~\ref{alg:one}). This solution works well for simple skeletons (the character skeleton used in the experiment has 32 joints and hierarchy layer < 10). However, for skeletons with many levels, the amount of computation per thread will be heavy, resulting in a poor efficiency.

\begin{algorithm}[t]
	\SetAlgoNoLine
	\For{all jointID in parallel
	}{
		$curParentID$ = $Parent(jointID)$\;
		\While{curParentID != invalid}{$M[jointID]$ = $M[curParentID] * M[jointID]$\;$curParentID = Parent(jointID)$\;}
	}
	
	\caption{\cite{Gateau2012monte}'s algorithm}
	\label{alg:one}
\end{algorithm}

KIYA KANDAR optimized Samuel Gateau's scheme in \cite{KIYA2024monte}. In the Hierarchy-Scan stage, threads are only allocated for leaf joints for calculation, and all parent nodes will be filled in during the recursive execution of leaf joints. This solution reduces the number of threads used for calculation, but for skeletons with many levels, the amount of calculation per thread is still very high.

In summary, although the existing work can optimize the calculation of skeletal animation through GPU in some limited circumstances (pre-baking/only targeting skeletons for characters with simple hierarchies), there is no GPU skeletal animation simulation solution that can retain the function of interpolating and blending animation data and can run efficiently on the skeleton tree of complex hierarchical characters.

\section{OVERVIEW}
In this article, we will introduce a GPU-driven skeletal animation system, which can compute animation simulation, Hierarchy-Scan, skinning and rendering in the GPU for characters with any complex skeletal hierarchical. Since animation simulation and skinning rendering currently have mature solutions that can be implemented in the GPU, this chapter will focus on describing how to efficiently implement Hierarchy-Scan in the GPU.

\subsection{Parallel bone tree update algorithm}
It can be found that the Hierarchy-Scan problem of skeleton tree can actually be regarded as a special prefix sum problem(Equation~\eqref{eqn:01}).
\begin{equation}
	\label{eqn:01}
	S_{i} = M_{i}  * M(Parent (i)) * M(Parent (Parent (i)))*…* M_{1} ,
\end{equation}
where $M_{i}$ is a given matrix,and its parent node ID is $Parent(i)$.

The parallel prefix sum problem is a problem with mature solutions. Hillis, W. Daniel, and Guy L. Steele proposed a parallel algorithm for prefix sum in a reduced way in \cite{hillis1986data}, and realized the parallel update of prefix sum with a complexity of $O(n\log_{2}{n})$. An idea based on binary balanced tree was proposed \cite{blelloch1990prefix}, further reducing the complexity of the algorithm to $O(n)$. \cite{harris2007parallel} analyzed how to implement these algorithms efficiently on GPU .

Based on the idea of prefix sum reduction and Mark Harris's implementation in GPU, we designed a parallel skeleton tree update algorithm(Algorithm~\ref{alg:two}).

\begin{figure}[htbp]
	\centering
	\includegraphics[width=0.5\textwidth]{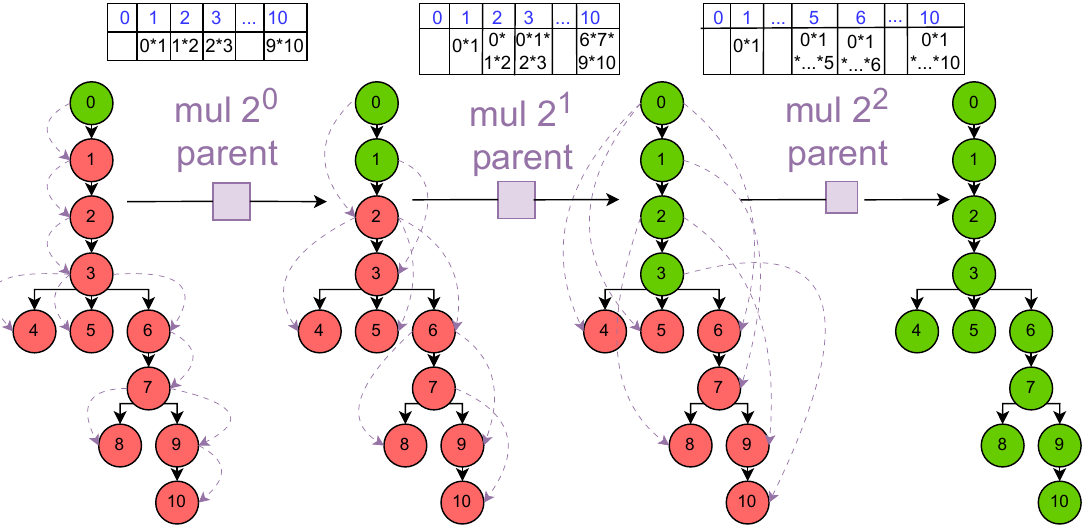}
	\caption{Parallel bone tree scan}
	\label{fig:four}
\end{figure}

\begin{algorithm}[ht]
	\SetAlgoNoLine
	\For{d = 1 to $\log_{2}{n}$
	}{
		\For{all jointID in parallel
		}{
			\If{MultiParent(jointID,d) !=invalid}{	M[jointID] = M[MultiParent(jointID,d)] * M[jointID]\;}
		}
	}
	\caption{parallel skeleton tree update algorithm}
	\label{alg:two}
\end{algorithm}

\begin{equation}[ht]
	\label{eqn:02}
	MultiParent(i,j) = \underset{j}{\underbrace{Parent(...Parent(Parent(i)))} }
\end{equation}
where $MultiParent(i, j)$to represent the $j-th$ level parent node of the $i-th~node$(Figure~\ref{fig:four}).

As Algorithm \ref{alg:two} states, In each operation, we multiply each node of the skeleton tree by its $pow(2,n)$ layer parent node. After $\log_{2}{n}$ operations, each node will obtain the calculation result from the root node to that node.
In this solution, each node on the skeleton tree will be assigned a GPU thread for calculation, and the final amount of computation for each thread is $O(\log_{2}{n})$. Compared with the traditional solution where the amount of computation for each thread is $O(n)$, this solution has higher efficiency when the number of skeleton levels is high. However, because the direct implementation of this solution requires $\log_{2}{n}$ barriers in the GPU. From our test results profiling, we can see that when the number of skeleton levels is not high, nearly $40\%$ of the performance overhead is spent on the GPU's memory barrier. Therefore, we need to improve this solution to reduce its memory barrier overhead in the GPU.

\subsection{Optimization for GPU hardware}
\subsubsection{Divide and conquer}
In the algorithm we proposed in $3.1$, the number of memory barriers in the GPU depends on the number of skeleton joints. Therefore, we designed a segmentation scheme to decompose the skeleton tree into multiple blocks with a number of nodes equal to $64$(Algorithm~\ref{alg:three}).
\begin{figure}[htbp]
	\centering
	\includegraphics[width=0.5\textwidth]{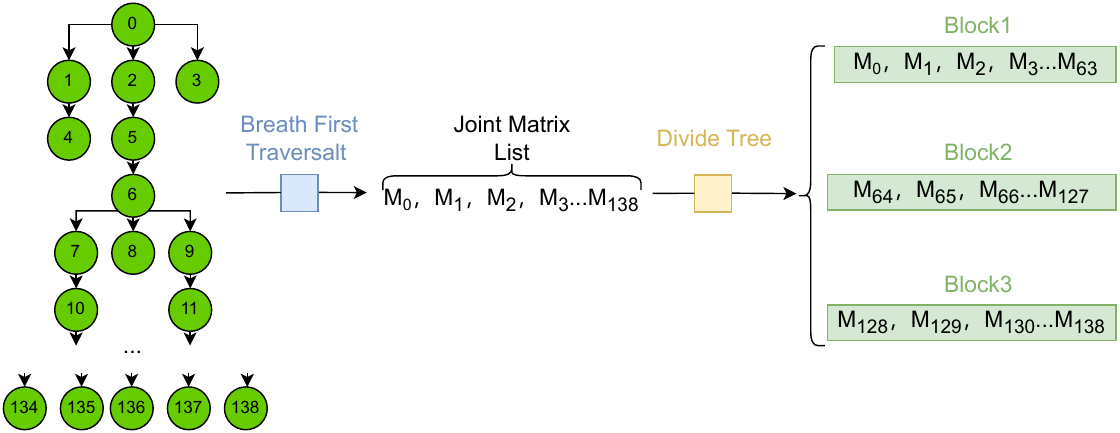}
	\caption{the processing of divided by block}
	\label{fig:five}
\end{figure}

As shown in the Figure~\ref{fig:five}, by traversing the skeleton tree in order, we can get the corresponding node list, and in this list, we can ensure that the parent node of any tree node must have a smaller sequence number in the list than the node. In this way, we can decompose all skeleton tree joints into blocks by node index. Based on grouping, we propose an optimization scheme, in which the amount of computation per thread is $6+n/64$. Compared with the $\log_{2}{n}$ operations of $3.1$, the amount of computation per thread of this scheme is slightly higher, but due to the reduction of the number of barriers, the overall performance is better than the scheme of $3.1$.

\begin{algorithm}[ht]
	\SetAlgoNoLine
	\For{d = 1 to 6
	}{
		\For{all jointID in parallel
		}{
			\If{MultiParent(jointID,d) !=invalid}{	M[jointID] = M[MultiParent(jointID,d)] * M[jointID]\;}
		}
	}
	\For{all jointID in parallel
	}{
	curParentID = MaxParentOutBlock (jointID)\;
	\While{ curParentID != invalid}{M[jointID] = M[curParentID] * M[jointID]\;
	curParentID = MaxParentOutBlock (jointID)\;}
	}
	\caption{divided by block}
	\label{alg:three}
\end{algorithm}

where $MaxParentOutBlock(i)$ represents the parent node with the largest ID that is not in the same block of the $i-th$ joint of the skeleton.

After group optimization, although the overhead of memory barrier is reduced, we found that the overhead of memory barrier still accounts for a large proportion when the number of skeleton levels is not high. Therefore, it is still necessary to continue to optimize the algorithm to further reduce the number of synchronizations.

\subsubsection{State compression}
In $3.2.1$, each block uses $6$ synchronization instructions when calculating internal matrix multiplication. However, each synchronization only calculates one matrix operation. This will seriously block the operation of the GPU thread. Therefore, we optimized the internal specification of each block.

\begin{algorithm}[ht]
	\SetAlgoNoLine
	\For{all jointID in parallel
	}{
		curParentID = Parent (jointID)\;
		\For{d = 1 to 7 
		}{
			\If{ curParentID != invalid}{
					M[jointID] = M[curParentID] * M[jointID]\;
					curParentID = Parent (curParentID)\;
					
					}
		}
		groupbarrier()\;
		curOutParentID = MultiParent(jointID,8)\;
		\For{d = 1 to 7 }{
			\If{ curParentID != invalid}{
				M[jointID] = M[curParentID] * M[jointID]\;
				curParentID = MultiParent(curParentID,8)\;
			}
		}
		groupbarrier()\;
		curParentID = MaxParentOutBlock (jointID)\;
		\While{ curParentID != invalid}{M[jointID] = M[curParentID] * M[jointID]\;
		curParentID = MaxParentOutBlock (jointID)\;}
	}
	\caption{State compression}
	\label{alg:four}
\end{algorithm}

As shown in the code, each node is allocated to multiply the parent node's matrix $8$ times continuously and then synchronized. In this way, only $16$ matrix multiplications and two barriers are required to complete all prefix operations in a block. At this time, the amount of calculation for each thread is $14+n/64$. Although each thread performs $8$ more matrix operations, the performance is greatly improved because four memory barriers are reduced. In the end, we adopted this solution as the final solution to resolve Hierarchy-Scan problem.

\section{Experiment}
We developed an experimental environment using $DirectX~12$ and implemented a complete GPU-based simulation and rendering process for large-scale skeletal animation(Table~\ref{tab:one}).
\begin{table}[h]
	\caption{Simulation Configuration}
	\label{tab:one}
	\begin{minipage}{\columnwidth}
		\begin{center}
			\begin{tabular}{ll}
				\toprule
				PARAMETER   & VALUE\\ \midrule
				Skeleton Charactor Number     & 10000\\
				Skeleton Joint Number  & 300*10000m\\
				Skeleton Hierarchy layer     & 15-120\\
				Mesh Face Number    & 1000-3000\\
				Graphic Api   & DirectX12\\
				Graphic Card       & RTX3080\\
				
				\bottomrule
			\end{tabular}
		\end{center}
		\bigskip\centering
	\end{minipage}
\end{table}%

\begin{figure}[htbp]
	\centering
	\includegraphics[width=0.5\textwidth]{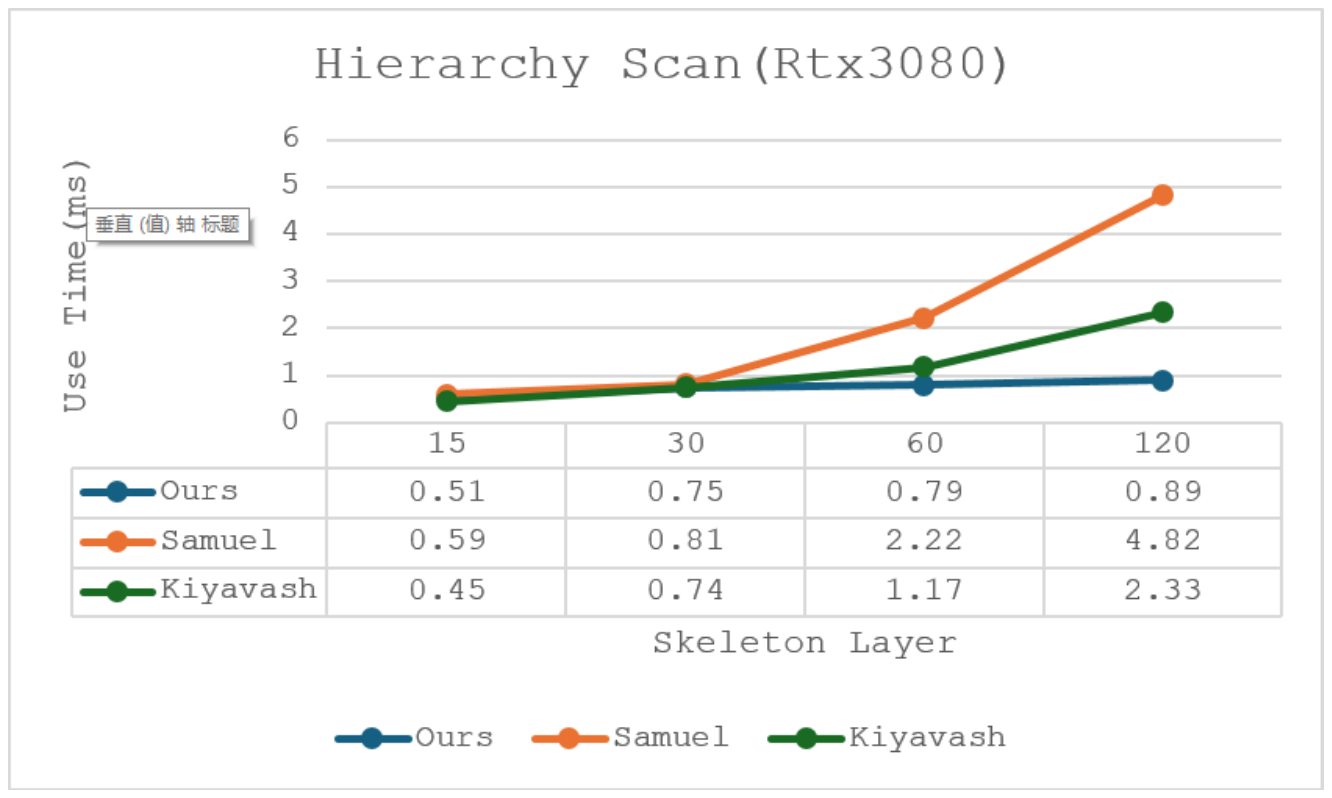}
	\caption{Performance comparison}
	\label{fig:sev}
\end{figure}

\begin{figure}[htbp]
	\centering
	\includegraphics[width=0.5\textwidth]{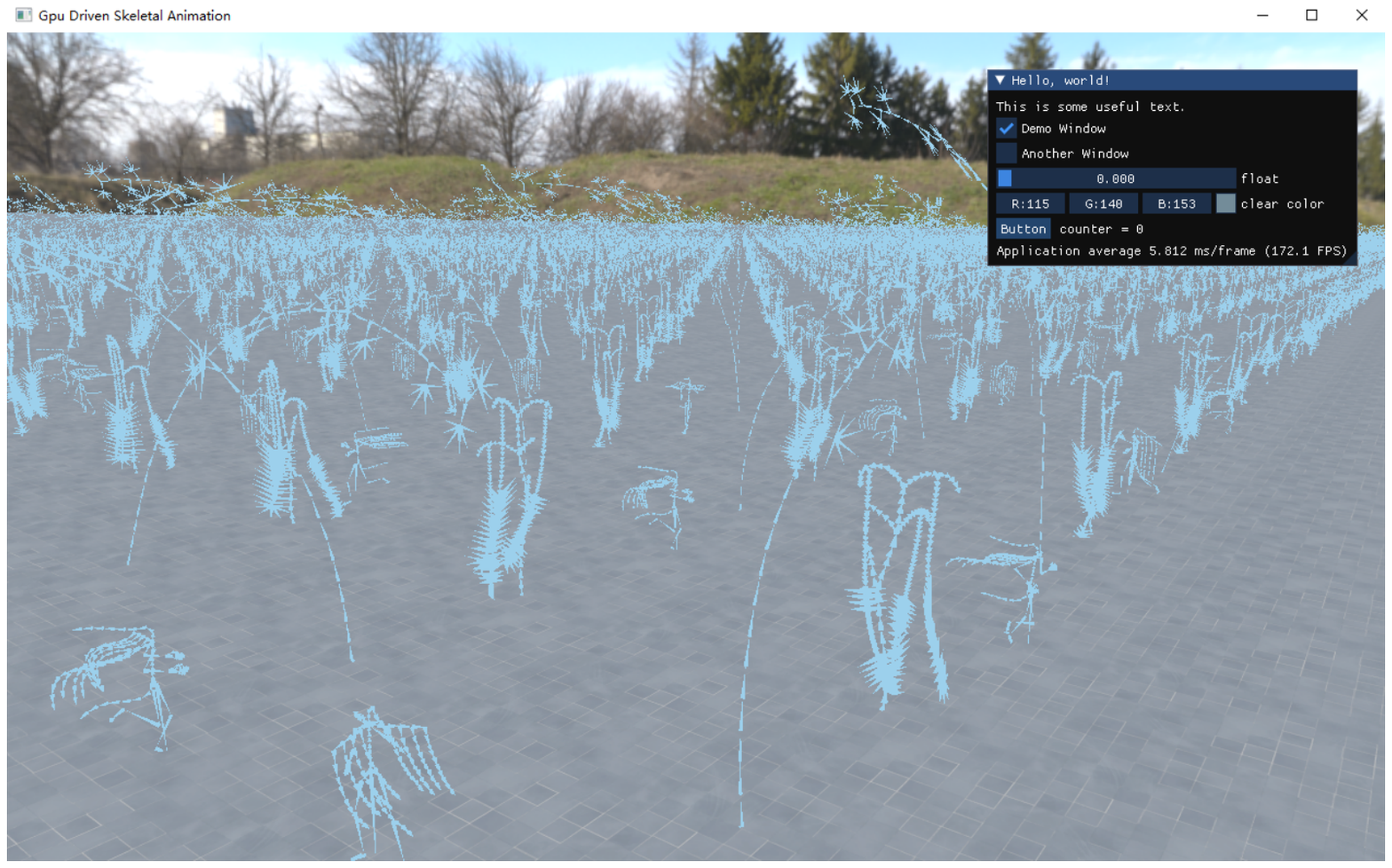}
	\caption{rendering result}
	\label{fig:six}
\end{figure}

To facilitate testing, we also implemented the solutions of \cite{Gateau2012monte} and \cite{KIYA2024monte} in the Hierarchy-Scan stage for comparison with our solution. The comparative experiment uses $10,000$ skeleton models for GPU animation simulation performance testing each time. The number of joints in each model skeleton is about $300$, that is, the number of skeleton joints involved in each animation operation is $300,0000$. In order to test the impact of the skeleton hierarchy on performance, the average hierarchy depth of the test model skeleton is gradually increased from $15$ layers to $120$ layers.

Figure~\ref{fig:sev} shows the performance comparison of our solution with \cite{Gateau2012monte} and \cite{KIYA2024monte} solutions at different levels of the Hierarchy Scan process when we use RTX3080 GPU through nsight profiling test. It can be seen that when the number of bone levels is small, the performance of our solution is close to that of \cite{KIYA2024monte} solution and is better than \cite{Gateau2012monte}. When the number of bone levels $>30$, the performance of our solution is significantly better than \cite{Gateau2012monte} and \cite{KIYA2024monte}.

In summary, the GPU-based skeletal animation simulation solution we proposed can maintain efficient operation on the skeleton tree of characters with complex skeleton hierarchies while retaining the function of interpolating and blending animation data. It is a more general solution.

\section{Conclusions}

In this paper, we proposed a GPU-based skeletal animation simulation solution. On the basis of traditional solutions, we implemented a solution based on parallel prefix tree updates to efficiently implement the hierarchical scan process of the skeleton tree on the GPU, solving the problem that the traditional solution has a serious performance degradation when dealing with too high a skeleton tree hierarchy. Our solution and experimental results can provide some new optimization ideas for the performance optimization of large-scale skeletal animation simulation of large-scale video games in the future. At the same time, our work can also provide some references for the design of large-scale cluster design tools for game engines in the future.

\bibliographystyle{ACM-Reference-Format}
\bibliography{draft-bibliography}


\begin{thebibliography}{7}


\ifx \showCODEN    \undefined \def \showCODEN     #1{\unskip}     \fi
\ifx \showISBNx    \undefined \def \showISBNx     #1{\unskip}     \fi
\ifx \showISBNxiii \undefined \def \showISBNxiii  #1{\unskip}     \fi
\ifx \showISSN     \undefined \def \showISSN      #1{\unskip}     \fi
\ifx \showLCCN     \undefined \def \showLCCN      #1{\unskip}     \fi
\ifx \shownote     \undefined \def \shownote      #1{#1}          \fi
\ifx \showarticletitle \undefined \def \showarticletitle #1{#1}   \fi
\ifx \showURL      \undefined \def \showURL       {\relax}        \fi
\providecommand\bibfield[2]{#2}
\providecommand\bibinfo[2]{#2}
\providecommand\natexlab[1]{#1}
\providecommand\showeprint[2][]{arXiv:#2}

\bibitem[Blelloch(1990)]%
        {blelloch1990prefix}
\bibfield{author}{\bibinfo{person}{Guy~E Blelloch}.}
  \bibinfo{year}{1990}\natexlab{}.
\newblock \showarticletitle{Prefix sums and their applications}.
\newblock  (\bibinfo{year}{1990}).
\newblock


\bibitem[Dudash(2007)]%
        {dudash2007skinned}
\bibfield{author}{\bibinfo{person}{Bryan Dudash}.}
  \bibinfo{year}{2007}\natexlab{}.
\newblock \showarticletitle{Skinned instancing}.
\newblock \bibinfo{journal}{\emph{NVidia white paper}} (\bibinfo{year}{2007}).
\newblock


\bibitem[Gateau(2012)]%
        {Gateau2012monte}
\bibfield{author}{\bibinfo{person}{Samuel Gateau}.}
  \bibinfo{year}{2012}\natexlab{}.
\newblock \showarticletitle{Massive Crowds on GPU,Siggraph 2012}.
\newblock \bibinfo{journal}{\emph{GTC 2012}} (\bibinfo{year}{2012}).
\newblock


\bibitem[Harris(2007)]%
        {harris2007parallel}
\bibfield{author}{\bibinfo{person}{M Harris}.} \bibinfo{year}{2007}\natexlab{}.
\newblock \showarticletitle{Parallel prefix sum (scan) with cuda}.
\newblock \bibinfo{journal}{\emph{GPU Gems}}  \bibinfo{volume}{3}
  (\bibinfo{year}{2007}).
\newblock


\bibitem[Hillis and Steele~Jr(1986)]%
        {hillis1986data}
\bibfield{author}{\bibinfo{person}{W~Daniel Hillis} {and}
  \bibinfo{person}{Guy~L Steele~Jr}.} \bibinfo{year}{1986}\natexlab{}.
\newblock \showarticletitle{Data parallel algorithms}.
\newblock \bibinfo{journal}{\emph{Commun. ACM}} \bibinfo{volume}{29},
  \bibinfo{number}{12} (\bibinfo{year}{1986}), \bibinfo{pages}{1170--1183}.
\newblock


\bibitem[KANDAR(2024)]%
        {KIYA2024monte}
\bibfield{author}{\bibinfo{person}{KIYA KANDAR}.}
  \bibinfo{year}{2024}\natexlab{}.
\newblock \showarticletitle{Large Scale GPU-Based Skinning for Vegetation in
  Alan Wake 2}.
\newblock \bibinfo{journal}{\emph{GDC 2024}} (\bibinfo{year}{2024}).
\newblock


\bibitem[Shopf et~al\mbox{.}(2008)]%
        {shopf2008march}
\bibfield{author}{\bibinfo{person}{Jeremy Shopf}, \bibinfo{person}{Joshua
  Barczak}, \bibinfo{person}{Christopher Oat}, {and} \bibinfo{person}{Natalya
  Tatarchuk}.} \bibinfo{year}{2008}\natexlab{}.
\newblock \showarticletitle{March of the froblins: simulation and rendering
  massive crowds of intelligent and detailed creatures on gpu}.
\newblock In \bibinfo{booktitle}{\emph{ACM SIGGRAPH 2008 Games}}.
  \bibinfo{pages}{52--101}.
\newblock


\end{thebibliography}

\end{document}